# Can the structure of amorphous indium gallium zinc oxide be described in terms of a few polyhedral motifs?

*Divya[a], R. Prasad[b], Deepak[a,c]*

[a] *Department of Materials Science and Engineering, Indian Institute of Technology, Kanpur, 208016, India*

[b] *Department of Physics, Indian Institute of Technology, Kanpur, 208016, India*

[c] *Samtel Center for Display Technologies, Indian Institute of Technology, Kanpur, 208016, India*

**Abstract:** The coordination polyhedra around the cations are the building blocks of ionic solids. In context of amorphous InGaZn oxide (a-IGZO), even though the coordination polyhedra are irregularly arranged, it will be beneficial to identify them, especially to investigate properties that emerge from short range or local interactions in the amorphous oxides. Accordingly, in this work, we address the questions, (a) is it possible to classify all the polyhedra that occur in a-IGZO into only a few distinct groups, and find their relative percentages of occurrence, so that commonalities can be identified and working with them becomes easier? and (b) are these the same polyhedral motifs as those observed in the crystalline indium gallium zinc oxide (c-IGZO) or other related crystalline oxides of indium, gallium and zinc? Therefore, in this first principles based study, a large number (ten) of equivalent samples of a-IGZO were prepared by ab initio melt-and-quench molecular dynamics, so that several distinct samples of the amorphous landscape are obtained corresponding to local minima in energy. The combination of all these structures thus obtained is a better representation of a real a-IGZO sample, rather than that obtained through only one or two simulated samples. For the ten samples containing 360 cations, we propose a simpler and more accurate method for determining the coordination number of each polyhedron, which was verified by charge density plots. Based on a method of comparing bond angles between metal and oxygen atoms, the identified polyhedra were matched to the polyhedral motifs present in the related crystalline systems, such as, $InGaZnO_4$, $In_2O_3$, $Ga_2O_3$ and ZnO. Accordingly, we find, the a-IGZO primarily consists of the following polyhedra: a tetrahedron from space group 199 and an octahedron from space group 206 of $In_2O_3$; a tetrahedron from space group 12 and an octahedron from space group 167 of $Ga_2O_3$; a tetrahedron from space group 186 of ZnO; zinc and gallium trigonal bipyramids from c-IGZO; and one zinc 4-fold, one zinc 5-fold and one indium 5-fold coordination polyhedra that occur only in the amorphous phase. Thus, we are able to reduce the description of structure from 360 to 10 groups of polyhedra. The benefits of this identification could be enormous. For example, now it may be possible to identify equivalent defect sites.

## I. INTRODUCTION

Amorphous indium gallium zinc oxide (a-IGZO) system is, at present, a popular choice for amorphous transparent oxide semiconductor[1-4], especially as an active channel layer for such electronic devices as active-matrix organic light emitting diodes (AMOLEDs)[5]. It exhibits tunable conductivity and high mobility (> 10 cm$^2$/V-sec) of similar magnitude as the c-IGZO system. This is attributed to the fact that in a-IGZO the conduction band minimum is mainly composed of metal *4s* or *5s* (*4*s for zinc and gallium; and 5s for indium) orbitals that are insensitive to disorder of the amorphous phase. In comparison a:Si-H exhibits modest values of mobility (1-2 cm$^2$/V-sec). Moreover, a-IGZO is deposited at a low temperature, is transparent because of its large bandgap and requires low cost for processing. Nevertheless, in the a-IGZO system, other issues, such as, subgap states near valence band maximum and negative bias illumination stress instability, emerge that compromise device performance [6-8].

In order to investigate the effects mentioned above and other properties relevant to device engineering, various first principles studies of the a-IGZO and c-IGZO systems have been undertaken[7-23] and role of oxygen defects has also been examined in much detail [8, 9, 14, 16, 19, 22]. Specifically, several studies regarding the amorphous structure of a-IGZO were also undertaken. In one of the earliest studies by means of x-ray absorption fine structure (XAFS) measurements and ab initio calculations it was found that the local coordination in the amorphous phase up till the nearest neighbor remained quite similar to that in the crystalline phase. However, behavior beyond the nearest neighbor varied for both of them[10]. In another study, these results were further confirmed, that is, the local oxygen-metal coordination from the crystalline state was preserved [24]. Moreover, in a study on a related system of metal oxides (In-X-O where X maybe Zn, Ga, Sn or Ge), it was found that a large proportion of indium atoms were five-fold coordinated[25]. This study further confirmed that the short range interactions that can be described in terms of coordination polyhedra of metals remains largely unchanged from the crystalline phase. However, these conclusions are based on only a few simulated samples, whereas a real sample would exhibit structures due to many structures obtained by cooling into a local energy minimum.

However, the local coordination structure for every atom varies even for the same elemental species in an amorphous system. Therefore, it becomes difficult to identify which sites are similar and which are different. This increases the complexities of problems like substitutional doping or creating vacancies when compared to a crystalline system where only a few distinguishable sites exist[7, 11-13]. Consequently, every site needs to be treated separately and it is difficult to predict apriori the contribution of each site to the electronic structure. Therefore, it is necessary to draw generalizations so that predictions can be made as to which sites are similar. This problem is dealt with by characterizing every cationic site by an associated coordination polyhedron. The objective of this work is to describe the amorphous IGZO structure in terms of its irregularly arranged coordination polyhedra. Simplistically, a polyhedra would be associated with each cation. Hence, the structure could be described by providing details of each polyhedra. But since the data would be enormous, it would be of little utility. Hence, in this work, we attempt to classify these polyhedra into groups so that description of the structure becomes meaningful. In order to do so, these polyhedra are compared with the cationic polyhedra occurring in crystalline system of related oxides, which are, c-IGZO, crystalline indium oxides, crystalline gallium oxides and crystalline zinc oxides. The question we answer is if the polyhedra in the amorphous state originate from the crystalline phases and if not, can they still be grouped. If this is so, then the amorphous structure can be merely described as an irregularly stacked polyhedra network composed of only a few types of polyhedra.

To address these questions, we prepare a large number of a-IGZO samples, identify their coordination polyhedra and compare them with those present in related crystalline phases.

## II. CALCULATION DETAILS

All the first principles calculations are performed using Vienna Ab initio Simulation Package (VASP)[26-29]. The generalized gradient approximation Perdew-Wang 1991 (GGA-PW91)[30, 31] was used for exchange correlation potential with projector augmented wave potentials (PAW)[32, 33] for ionic potentials. There are ten equivalent a-IGZO samples, each composed of 84 atoms, that is, twelve formula units of InGaZnO$_4$. In the subsequent sections, we have labeled the 48 oxygen atoms in each sample from O1 to O48. Similarly, the 12 indium, gallium and zinc atoms are labeled In1 to In12, Ga1 to Ga12 and Zn1 to Zn12 respectively. In each sample there are 13 valence electrons for In ($4d^{10}\ 5s^2\ 4p^1$), 13 for Ga ($3d^{10}\ 4s^2\ 4p^1$), 12 for Zn ($3d^{10}\ 4p^2$) and 6 for O ($2s^2\ 2p^4$).

Melt and quench ab initio molecular dynamics was performed on each a-IGZO sample using Nosé-Hoover thermostat[34, 35]. Thus, ten separate molecular dynamics simulations are performed, each using the number-volume-temperature (NVT) ensemble. The fictitious Nosé mass is determined using the maximum phonon frequency of ZnO [36]. Each of these ten a-IGZO samples are heated to 3000 K to remove structure memory effects and the cooling cycles are sufficiently long so that all the samples are equivalent.. Consequently a number of possible variations for local coordination for all the atomic species are acquired so that the results obtained are not merely the characteristic features of any one model but can be generalized to the real system. Thus, these samples are representative samples and real sample is some weighted average of all these samples. This is the reason for evaluating all the samples together, as further described in the results and discussion section.

The details regarding the cooling cycles and initial conditions for the molecular dynamics for the other samples are given in the supplementary information (Section 1). As an example, the initial structure for the first sample was the crystalline phase which was melted at 3000K for 10fs. Thereafter, the sample temperature was progressively reduced by a step of 500K and the temperature was held constant for 5 ps at each step. This process was followed up till 500K, where after the sample was cooled to 300k at the cooling rate of 0.1K/fs. At 300K, the temperature was again held constant for 5ps. Then, the sample was finally cooled to 0K at the rate of 0.1K/fs. These simulations were carried at an energy cutoff of 205 eV and a Gamma k-point mesh of size 1X1X1. Thereafter, volume and ionic relaxations are done on these a-IGZO samples till the forces are less than 0.01 eV/Å. The relaxation calculations are also carried out with a Gamma k-mesh of size 1X1X1 and energy cutoff of 400 eV. Much of the calculation methodology is similar to that given in a previous work [16]. The samples so obtained are deemed as the final samples and an examination and comparison of their structure is carried out in the following sections. All visualizations of structure including isolated coordination polyhedra and charge density plots are performed using the visualization software VESTA[37].

## III. RESULTS AND DISCUSSION

An amorphous structure is generated due to a local minima in energy. In the energy landscape many such minima exist. A real a-IGZO sample consist of a collection of structures corresponding to all such local minima. However, most studies with ab-initio simulated annealing derive their conclusions based on only 1-3 samples [16]. Therefore, we suggest that in order to elicit meaningful conclusions with respect to amorphous structures, a larger number of samples are needed. In this work, we generate ten samples. Each sample would cool into a structure corresponding to a different local minimum in energy. A real system would be represented as one having all these structures. Also, the large number of local structures generated in this work allow statistical analysis.

The structure of crystalline ionic solids can be described as a regular packing of cationic polyhedra. An amorphous structure would have disordered arrangement of polyhedra. But do these polyhedra in the amorphous structure closely resemble those found in the corresponding crystalline structures? This is the question we wish to address here. Therefore, the first step is to identify the polyhedra present in all the ten a-IGZO samples prepared in this study.

Each a-IGZO sample has 12 indium, gallium and zinc atoms each and each of these metal atoms are expected to bond with either four, five or six oxygen atoms based on the observations from related crystalline phases. Usually, the pair correlation function (PCF) is deployed to determine if a bond exists between the metal atom and surrounding oxygen atoms[16, 38]. This method is based on the choice of a cutoff distance for bonding, judged on the basis of the distribution represented by the PCF. However, the choice of this cutoff remains arbitrary. Whereas in the study by Noh et al [16], the cutoff radius for indium was 2.9 Å, the cutoff radius in this study was varied from 2.5 to 3.0 Å for a set of 36 indium polyhedra (from samples 2, 9 and 10, see supplementary information, section 1 and 2). As a result, the coordination numbers of several indium atoms changed and it was difficult to fix a cutoff that would give uniformly accurate coordination numbers for all the 36 indium atoms. Over only a range of 0.2 Å, the number of 5-fold polyhedra changed from 9 to 6 while the number of 6-fold polyhedra changed from 23 to 25 as the cutoff radius changed from 2.8 to 3.0 Å. Therefore, a choice of a rigid cutoff will necessarily result in erroneous conclusions due to over-bonding (implying additional oxygen atoms are included than actually bonded to the metal atom) or under-bonding. In the present study we wish to identify individual polyhedra for each cation and hence the PCF approach is not adequate to determine the coordinating oxygen atoms accurately. Thus, we propose an alternate approach for determination of coordination polyhedra in amorphous ionic systems.

## A. Alternate approach for determining coordination polyhedra

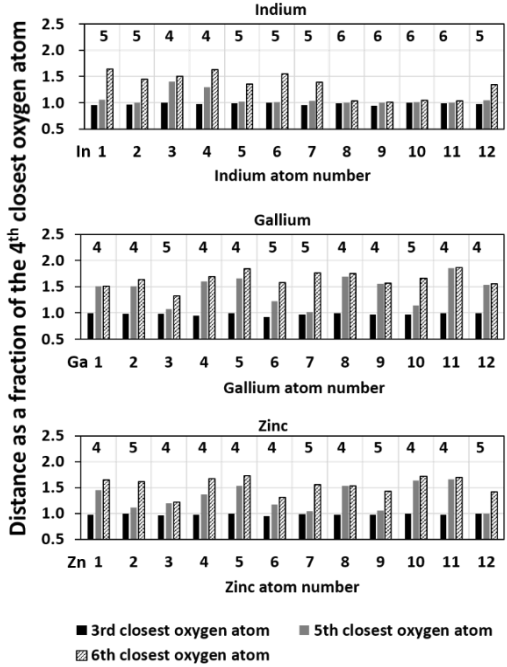

**Figure 1:** Distance of the third, fifth and sixth closest oxygen atom as a fraction of the distance of the fourth closest oxygen atom (sample number 2 – see supplementary information section 1). The coordination number for each cation is mentioned above the respective bar graph.

We examine one of the ten samples at a time. The indium, gallium and zinc atoms can be four-, five- or six-fold coordinated. Therefore, for each cation in that a-IGZO sample, we determine the distance of $5^{th}$ and $6^{th}$ nearest oxygen relative to the $4^{th}$ nearest oxygen atom. These relative positions as fractions are depicted in Figure 1, where as an additional reference, we have included the relative position of the $3^{rd}$ oxygen atom as well. In short, for

$$d_i = \frac{distance\ of\ the\ i^{th} nearest\ oxygen\ from\ a\ cation}{distance\ of\ the\ 4^{th}\ nearest\ oxygen\ from\ the\ same\ cation}, \qquad (1)$$

$d_3$, $d_5$ and $d_6$ are plotted in Figure 1. The $4^{th}$ nearest oxygen is used as a standard to be compared with, because 4-fold is assumed to be the start of coordination in 3-dimension for indium. Similar plots are prepared for the remaining samples also (not shown here).

We propose that the bonding and non-bonding neighbors can be distinguished through this plot, based on the idea that all bonded oxygen atoms should have bond length values in a narrow range. For every cation, $d_3$ is greater than 0.9, which means it differs by less than a step difference of 0.1 from $d_4$. The step difference between any two consecutive $d_i$'s should be small for bonding to exist. Hence, if we decide that the step difference should be no more than 0.1, it sets the benchmark that $d_5$ could be approximately 1.1 and $d_6$ approximately 1.2, for $5^{th}$ and $6^{th}$ oxygen atom to be bonded to the cation under consideration. To illustrate this, for the indium atom labeled '1' in Figure 1, note that the bond distances of $5^{th}$ and $6^{th}$ closest oxygen atoms are such that $5^{th}$ atom can be considered bonded ($d_5 \approx 1.1$) while the $6^{th}$ atom is located so far away ($d_6 \approx 1.6$). As a result, the coordination can be unambiguously indicated as being 5-fold. Accordingly, the coordination number is also included at the top in the bar graph of Figure 1. Coordination of all other indium atoms is assigned subsequently. However, this new approach requires a reliability test, which can be provided by the charge density plots.

The same indium atom, labeled In1 in Figure 2 has its six closest oxygen atoms labeled O11, O33, O40, O31, O10 and O19 arranged in increasing distance from the In1 atom. The charge density plots in Figures 2(a) to 2(d) are plotted for the isosurface value of 0.06 electrons/Bohr$^3$ where it is observed that the nature of bonding is partly covalent indicated by an overlap of electron clouds between indium and oxygen atoms. However, at the same isosurface level, an overlap between the metal and oxygen charge density is not observed for the fifth oxygen atom, though the electron cloud on O10 is clearly distorted (Figure 2(e)). Therefore, as the isosurface value is reduced to 0.045 electrons/Bohr$^3$, the distortion in Figure 2(e) morphs into an overlap between the charge densities of In1 and O10 atoms in Figure 2(f), indicating bonding of the fifth oxygen. However, no overlap (or distortion) is observed between In1 and O19 that would indicate that any bond exists atoms, even when the isosurface value is reduced to 0.03 electrons/Bohr$^3$ in Figure 2(g). Therefore, the indium atom in Figure 2 is 5-fold coordinated which supports the conclusion drawn from Figure 1. Thus, the tedious examination of charge density plots as described here gives us a reliable method to determine if bonds exists between a metal atom and its surrounding oxygen neighbors. The simpler method proposed previously gives the same result.

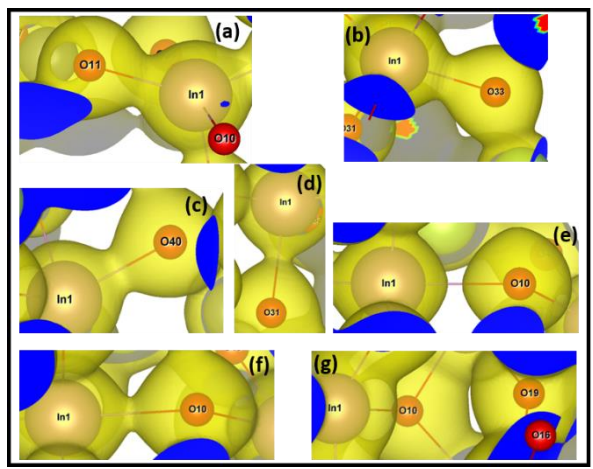

**Figure 2:** Charge density isosurfaces for an indium cation and its six oxygen neighbors. Figures (a)-(g) depict the various oxygen neighbors (O11, O33, O40, O31, O10 and O19) located at successively farther distance from the indium atom. Figures (a) to (e) are plotted for the isosurface value of 0.06 electrons/Bohr$^3$. Figure (f) and (g) are plotted at the isosurface value of 0.045 electrons/Bohr$^3$ and 0.03 electrons/Bohr$^3$, respectively. Except O19, all other oxygen atoms are bonded to the In1 atom; therefore, this indium atom is five-fold coordinated.

Similarly, we have determined the coordination number for the gallium atoms indicated in Figure 1. However, an ambiguity arises for the gallium atoms labeled 3 and 6. In case of Ga3, $d_5$ is such that 5$^{th}$ oxygen is clearly bonded, but $d_6$ is not so large that its bonding with cation can be unequivocally rejected. Similarly, for Ga6, height steps in Figure 1 are too equally spaced to reject outright 4-, 5- or 6-fold coordination. In the same way while the coordination of most of the zinc atoms could be determined, ambiguity arose for zinc atoms, also labeled 3 and 6; in the case of Zn3, $d_5$ and $d_6$ are not so far from the benchmark values and for zinc 6, the steps are again equally spaced. Thus, we conclude that these two types of conditions lead to ambiguity, while in majority of cases the coordination number is easily assigned.

In any case, for rigorous testing of the methodology, bonding in all 36 atoms was determined by charge density plots and the conclusions drawn matched accurately with Figure 1 in all cases where unambiguous assignments could be made. The coordination for gallium atoms 3 and 6 and zinc atoms 3 and 6 was also determined through charge density plots, and thus included in above the bar graph in Figure 1. The coordination of atoms in the remaining nine a-IGZO samples was determined through fractional distance plots similar to Figure 1 and in all cases results were verified with the respective charge density plots.

In short, when several coordination polyhedra exist in a structure and polyhedra associated with individual cations are to be determined, a method similar to PCF would not be suitable. The obvious alternative is a charge density plot around each cation, visualized under several sections and charge isosurfaces, which would be enormously laborious. The method presented here accurately determines the polyhedra coordination, unambiguously in most cases. In a small number of cases, where ambiguity could exist, the conditions are clearly established, which allows examination of charge density plots only in a limited number of cases to resolve the ambiguity.

Based on this approach the average bond lengths for In-O, Ga-O and Zn-O bonds are observed to be 2.28 Å, 1.99 Å and 2.12 Å respectively. Whereas in another ab initio based study[16] using PCF approach, where bonding was decided based on the choice of a cutoff, the average bond lengths are smaller - 2.15 Å, 1.79 Å and 2.00 Å respectively. However, the method presented here is more direct and hence accurate.

### B.    Methodology for Comparing Polyhedra

Once the polyhedra present in all the ten a-IGZO are identified, they are compared with the various polyhedra identified in several crystalline phases. Those that do not resemble with their crystalline counterparts are still further compared for any commonalities amongst themselves.

In order to do so, the indium, gallium and zinc polyhedra obtained from the amorphous phase are compared with the polyhedra from the various crystalline systems of indium oxides [39], gallium oxides [40, 41], zinc oxides [42-44] and crystalline InGaZnO$_4$ [45, 46]. All these crystalline polyhedral templates are depicted in Figure 3.

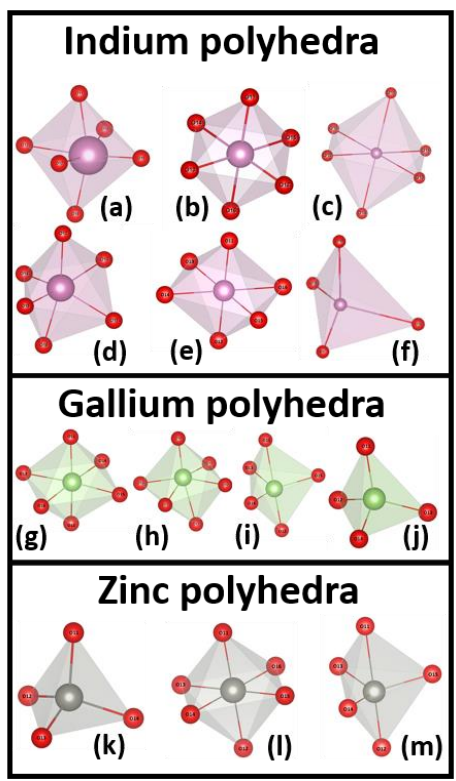

**Figure 3:** Crystalline polyhedral templates with which all the polyhedra in the amorphous system are compared. (a) octahedron in $In_2O_3$ – space group 167 (SG 167), (b) octahedron in $In_2O_3$ - space group 199 (SG 199), (c) one of the two octahedron in $In_2O_3$ of space group 206 (SG 206-1), (d) the second octahedron in $In_2O_3$ of space group 206 (SG 206-2), (e) indium octahedron from c-IGZO ($InGaZnO_4$), (f) tetrahedron in $In_2O_3$– space group 199 (SG 199), (g) octahedron in $Ga_2O_3$– space group 12 (SG 12), (h) octahedron in $Ga_2O_3$– space group 167 (SG 167), (i) gallium trigonal bipyramid from c-IGZO ($InGaZnO_4$), (j) tetrahedron in $Ga_2O_3$– space group 12 (SG 12), (k) tetrahedron in ZnO - space group 186 (SG 186), (l) octahedron in ZnO– space group 205 (SG 205), (m) zinc trigonal bipyramid from c-IGZO ($InGaZnO_4$) [Figures drawn with VESTA code [37]]

Comparison between two given polyhedra is usually done based on quantities like polyhedral volume[47], average bond length, distortion index[48], quadratic elongation[49], bond angle variance[49] and effective coordination number[49-51]; but most of these quantities are measures of distortion from the regular polyhedron of identical coordination number. However, in this study we are comparing polyhedra in a-IGZO with those in related crystalline oxides, which are not regular polyhedron. Since an exact match between two polyhedra in the amorphous phase and the crystalline phase will be extremely rare, therefore a new quantitative measure is to be devised. A closer examination of the crystalline structures in Figure 3 reveals that these polyhedra are distorted. Further, the bond lengths within a polyhedron with respect to the central cation show only small deviations, much less than those in bond angles. That is, significant distortion in a polyhedron is due to bond angles. Hence, the quantitative measure to be devised should be based on bond angles. The measure and its implementation are best illustrated through an example in which the polyhedron associated with atom labeled Ga2 in Figure 1 is compared with the polyhedron of crystalline phase shown in Figure 3(j).

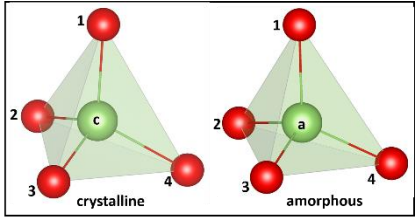

**Figure 4:** Bond angle comparison of two tetrahedra. Tetrahedron labeled *c* is obtained from crystalline $Ga_2O_3$ and tetrahedron labeled *a* is chosen from one of the amorphous samples.

The polyhedron in Figure 4 having central atom '*c*' is from crystalline phase (see Figure 3(j)), where the oxygen atoms labeled 1 through 4 are fixed. The polyhedron with central atom '*a*' is of a-IGZO (Ga2 in Figure 1), in which oxygen labels 1 through 4 are picked in one arbitrary order and changed subsequently, as described later. Nonetheless, this is one of the orientations in which bond angle comparisons were made. Then, the differences in bond angles *1-c-2* and *1-a-2*, *1-c-3* and *1-a-3*, and so on were calculated and tabulated in Table 1.

**Table 1:** Bond angles for the polyhedron *c* (crystalline) compared with bond angles for the polyhedron *a* (amorphous), where the orientation for the polyhedron *a* is picked according to Figure 4. For example, in 1-*-2, * stands for either *a* or *c* and it represents bond angle *1-a-2* or *1-c-2*, respectively.

| Index, $i$ | Bond angle labels | Bond angles for polyhedron $c$ (º) | Bond angles for polyhedron $a$ (º) | Difference in bond angles (º) |
|---|---|---|---|---|
| *1* | *1-\*-2* | 105.8 | 108.8 | 3.0 |
| *2* | *1-\*-3* | 119.1 | 120.4 | 1.3 |
| *3* | *1-\*-4* | 105.8 | 99.7 | 6.1 |
| *4* | *2-\*-3* | 107.6 | 107.5 | 0.1 |
| *5* | *2-\*-4* | 110.9 | 114.9 | 4.0 |
| *6* | *3-\*-4* | 107.6 | 105.9 | 1.7 |

Thereafter, the root mean square (RMS) of differences calculated in the table above is evaluated according to.

$$RMSD = RMS\ of\ differences = \sqrt{\sum_{i=1}^{n} \frac{|a_i^{BoA} - c_i^{BoA}|^2}{n}} \qquad (2)$$

where, $a_i^{BoA}$ and $c_i^{BoA}$ are $i^{th}$ bond angle in the polyhedron from the amorphous and crystalline phase, respectively and *n* is the number of distinct bond angles for a polyhedron (n = 6, 10 and 15 for 4-, 5- and 6-fold coordinated polyhedra, respectively) The value of the RMSD for the values listed in Table 1 is 3.

For comparison of bond angles, the labels *1* through *4* on *a* were changed while keeping labels *1* through *4* on *c* fixed, obtaining a different orientation for bond angle comparison. For a tetrahedron there are 24 (4!) unique ways to compare the bond angles. The RMS of differences in bond angles was recalculated for each of these new orientations. The minimum RMSD value (out of twenty-four orientations for a tetrahedron), named RMSD$_{min}$, reflects the best orientation for comparing polyhedra *a* and *c*.

It still remains to be decided as to what RMSD$_{min}$ value provides a good match between two polyhedra being compared. Therefore, a suitable cutoff for the RMSD$_{min}$ value is determined for comparing six-fold, five-fold and four-fold coordinated polyhedra. We have investigated a large number of cut-off values. Based on it, we demonstrate that a cut-off of 15º is adequate, as evidenced in Figure 5. The polyhedra from crystalline phases in Figure 5 are those which eventually have been matched to amorphous structures, as explained in a greater detail in the next section. Here, we only show the worst matched polyhedra from the amorphous phase for a cut-off value of 15º. For example, consider In tetrahedron in Figure 5. Many 4-fold coordinated polyhedra in a-IGZO matched space group (SG 199) tetrahedron from a crystalline phase with RMSD$_{min}$ ≤ 15º. The a-IGZO tetrahedron compared with it in Figure 5 is the poorest matched one with RMSD$_{min}$ of 13º; all other matched tetrahedra in this group had lower RMSD$_{min}$ values and those not matched, a value greater than 15º. Clearly, a visual examination establishes that the cut-off criterion used here is adequate and any two polyhedra that show RMSD$_{min}$ value less than 15º can be considered alike.

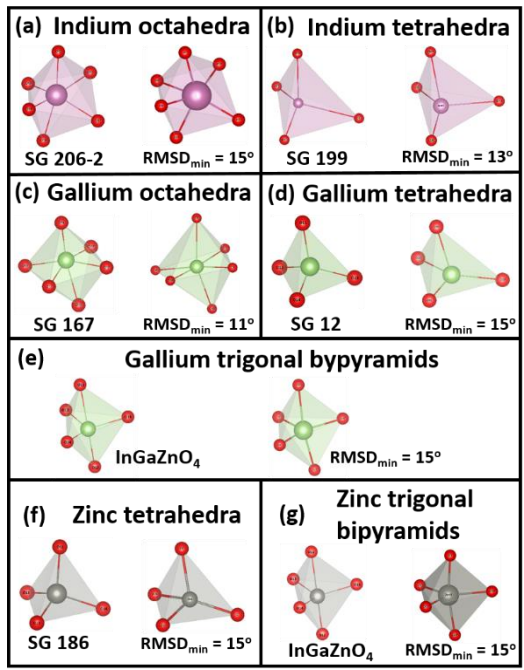

**Figure 5:** Crystalline polyhedra on the left in each block is followed by the corresponding amorphous polyhedra on the right. The amorphous polyhedra differ from the crystalline by an RMSD$_{min}$ ≤ 15º. (a) indium SG 206-2; (b) indium SG 199; (c) gallium SG 167; (d) gallium SG 12; (e) gallium InGaZnO$_4$; (f) zinc SG 186 and (g) zinc InGaZnO$_4$.

In all, there are 360 cations, and as many polyhedra, from the ten samples. Now we are in a position to categorize them in a smaller set of building blocks of the amorphous structure.

### C.     Classification of polyhedra obtained from the amorphous samples

Based on the analysis outlined in the preceding section the percentages of occurrences of various crystalline polyhedra is determined (Figure 6). Starting with 120 indium polyhedra, only 8 were 4-fold coordinated and of these, 6 (or 5% of total) could be matched to indium tetrahedron in In$_2$O$_3$ structure (SG 199). As for the remaining two, further grouping is also possible based on whether or not they compare between themselves. But since the number is so small, we ignore this question now (but discuss later in other cases such as Zn). Most of the In polyhedra were either 5- or 6-fold coordinated, 47 and 65, respectively. There are no 5-fold polyhedra identified from related crystalline structures. However, there are five indium octahedral motifs, as reported in Figure 3. Out of 65 six-fold coordinated polyhedra in a-IGZO, 58 (or 48% of total, see Figure 6(a)) could be matched to one of the indium octahedron in $In_2O_3$ structure (space group 206, or SG 206-2). Several 6-fold In polyhedra also matched the other octahedral motifs from the crystalline structures, but since these same also belonged to the SG 206-2 set, they were included in the largest group.

As for the 5-fold coordinated polyhedra, though there are no corresponding structures from crystalline phases, it may be possible to examine similarities, and hence make groupings, amongst themselves. To achieve this, the bond angles of these polyhedra are compared with each other. That is, each of 47 (out of 120, or 39%) of the indium polyhedra that have five-fold coordination is compared with the remaining 46. The polyhedron that matched the maximum number of others, based on the 15º cut-off criterion, is reported in Figure 6(a). Strikingly, in this way, 46 out of 47 (38% of total) 5-fold polyhedra could be matched amongst themselves; in other words the 5-fold polyhedra are almost all similar.

In principle, 120 In polyhedra could all have been structurally distinct in an amorphous phase. Important finding here is that, instead, a-IGZO can be represented by a small number of building blocks, octahedron similar to SG 206-2 (48% occurrence), 5-fold coordinated polyhedron present as one grouping (38% occurrence) and tetrahedron similar to SG 199 (5% occurrence).

Similarly, with respect to Ga polyhedra classification in Figure 6(b), the 120 seemingly distinct amorphous polyhedra correspond to mainly three polyhedral motifs, namely, tetrahedron SG 12 (Figure 3(j), 69% occurrence), trigonal bipyramid (Figure 3(i), 18% occurrence) and octahedron SG 167 (Figure 3(h), 4% occurrence). Almost all 4-fold coordinated polyhedra could be mapped to the tetrahedron from gallium oxide crystalline phase; only 4% remained unmatched and since the number is not too large, no comparison amongst them was made. Similarly, only 5% of 5-fold polyhedra remain unmatched, while all the 6-fold polyhedra could be mapped to the crystalline phase. Clearly, again 120 polyhedra could be described by only three types of polyhedra; the number which remains unmatched is small.

In one respect this study differs from another study done previously[10], where four-fold coordination was not observed and the Ga atoms were either 5-fold or 6-fold coordinated, as the running coordination number (RCN) did not show a sharply defined step for 4-fold coordination. However, the RCN is calculated from the PCF and therefore is subject to the same limitations as discussed earlier. A peak in the PCF will manifest as a step in the RCN plot and if the bond lengths for 4-fold and 5-fold coordination are close together then their steps will not be distinct. Therefore, if the margin of error for calculating the coordination number for each cation is small, then the way to ensure accuracy is to follow the methodology outlined here.

While in In and Ga, there were one or two dominant polyhedra, in the case of zinc, four major groups appear in similar proportions (Figure 6(c)). Among the 4-fold coordinated polyhedra (total 62%), 27% match with the zinc tetrahedron of the space group 186. Most of the remaining ones matched amongst themselves (32%, 4-fold) and a small number, 3% remained unmatched. Trigonal bipyramid from the c-IGZO phase could mapped to 12% of all Zn polyhedra, but a larger fraction, 22% of total, did not match the crystalline phase. However, amongst these, again most (19%, 5-fold) were similar and could be grouped together. The percentage of 6-fold polyhedra of Zn is small (4%). Consequently, four distinct sets are observed, two from crystalline phases – tetrahedron SG 186 and trigonal bipyramid and two that are only observed in the amorphous phase. At this stage we also note that there is another tetrahedron SG 216 from a ZnO crystalline phase[44], but it is almost identical to the tetrahedron considered here and hence not considered separately. Similar observation is made for octahedron SG 225 [44] which is equivalent to the octahedron SG 205 considered here.

In summary, the a-IGZO primarily consists of the following types of polyhedra: SG 199 tetrahedron and SG 206-2 octahedron from $In_2O_3$; SG 12 tetrahedron and SG 167 octahedron from $Ga_2O_3$; SG 186 tetrahedron from ZnO; zinc and gallium trigonal bipyramids from c-IGZO; and one zinc 4-fold, one zinc 5-fold and one indium 5-fold coordination polyhedra that occur only in the amorphous phase. Thus, we are able to reduce the description of structure from 360 to 10 groups of polyhedra.

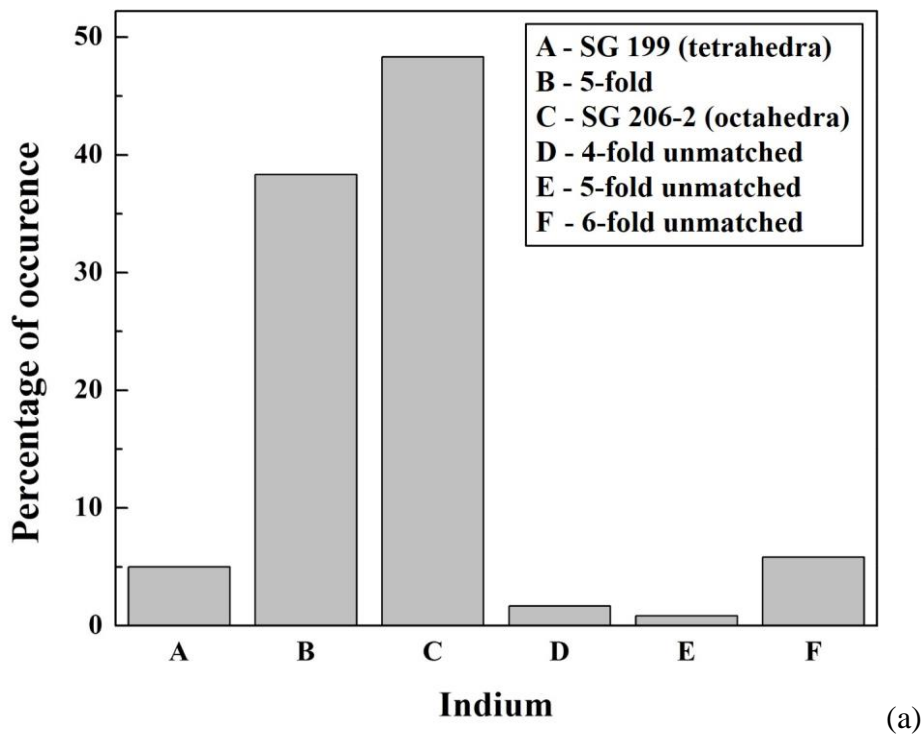
(a)

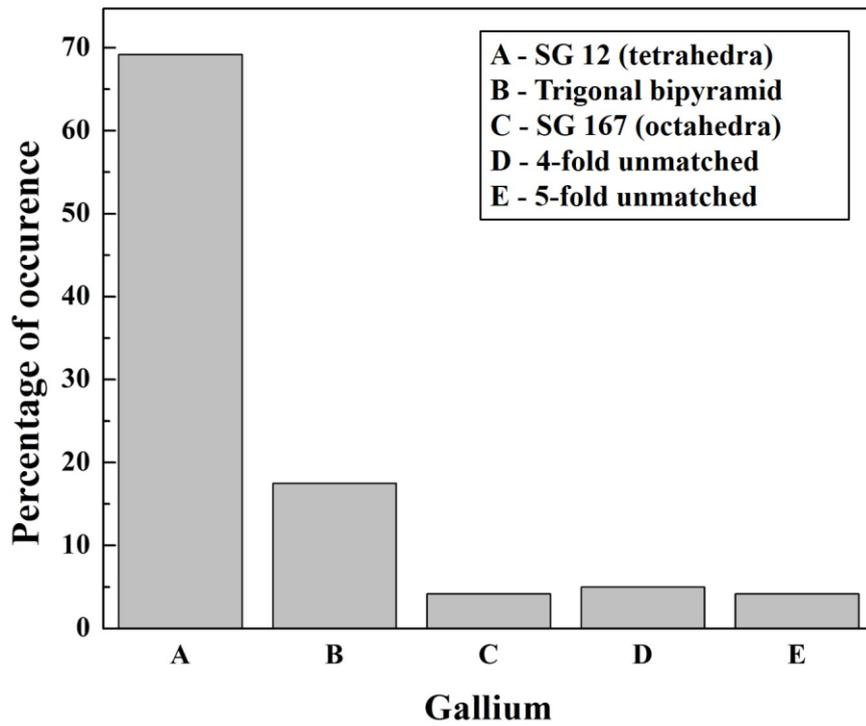

(b)

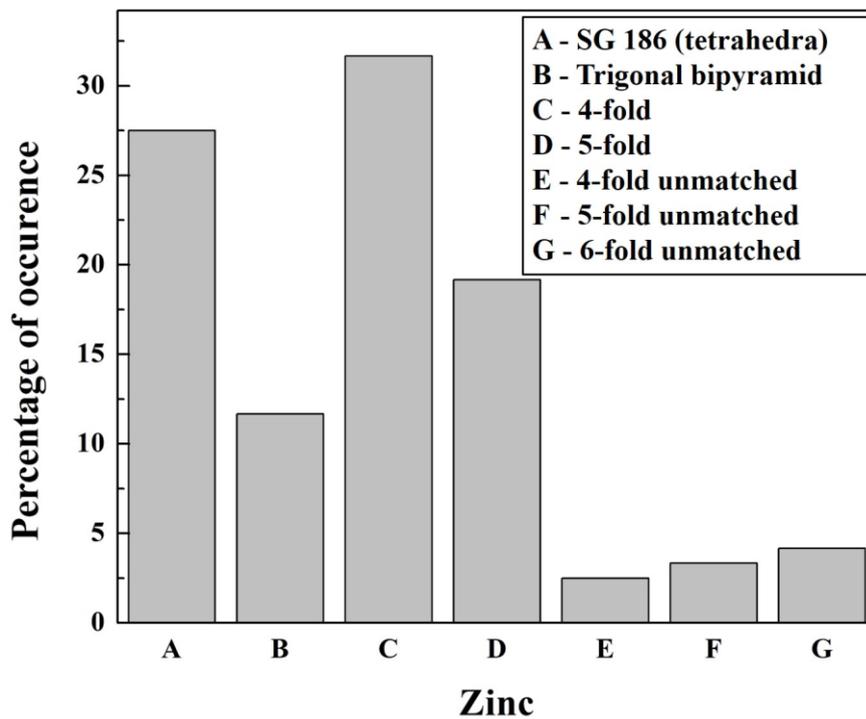

(c)

**Figure 6:** Types and percentages of various polyhedra occurring for (a) indium, (b) gallium and (c) zinc atoms in the a-IGZO.

## IV. CONCLUSION

In this study we have described the amorphous system of a multicomponent oxide (a-IGZO) by means of coordination polyhedra of the cations. This idea is a common one, but identification of coordination in polyhedra by such means as the pair correlation function suffers from infirmities. In this paper, we have addressed those and proposed a simple approach for identification of the coordination number of each cation. Furthermore, occasional ambiguities may arise, but importantly they are clearly identifiable in the proposed approach. These are then resolved by charge density plots.

These polyhedra are building blocks of both crystalline and amorphous phases in ionic systems. Whereas, in a crystalline structure the polyhedra are packed in a regular arrangement, an irregular network also exists in amorphous phases. Further, while polyhedra of same coordination number are mostly identical in crystalline systems, they are not in amorphous phases. Due to these variations found in polyhedra of same coordination in amorphous systems, it becomes difficult to identify commonalities in it. The existing approaches prevent us from finding any recurrent polyhedral motifs present in the amorphous system or if there exist equivalent sites based on local structure alone.

Therefore, we have proposed a method that allows apparently diverse polyhedra to be classified into only a few distinct groups derived from the local structure around each metal site. Based on the bond angles, this is done by comparing the polyhedra in a-IGZO with different polyhedral motifs from the various crystalline phases of indium oxide, gallium oxide, zinc oxide and indium gallium zinc oxide. Those polyhedra not closely matched to any polyhedra of the crystalline phases are then compared amongst themselves for further classification.

Accordingly, the In polyhedra in a-IGZO are adequately described by a single five-fold polyhedron, one tetrahedron and one octahedron both associated with crystalline $In_2O_3$. Similarly, all gallium polyhedra could be classified into a tetrahedron, trigonal bipyramid (5-fold) and an octahedron, all from crystalline phases. Four motifs could describe almost all of the zinc polyhedra. In short, we have identified 10 types of polyhedra in a-IGZO, along with their relative occurrence, which could be used to describe the a-IGZO structure.

The benefits of this identification could be enormous. For example, now it may be possible to identify equivalent defect sites, which can be used for doping at selective locations.

## ACKNOWLEGEMENTS

This work was supported by the Department of Science and Technology, New Delhi (India) through project SR/S2/CMP-0098/2010. Ms. Divya, a JRF scholar, was financially supported by the Council of Scientific & Industrial Research (CSIR), India.